\documentclass[pra,twocolumn,showpacs]{revtex4}%
\usepackage{amssymb}
\usepackage{amsfonts}
\usepackage{amsmath}
\usepackage{graphicx}
\usepackage{bm}
\usepackage{soul}
\usepackage{color}%
\providecommand{\U}[1]{\protect\rule{.1in}{.1in}}
\soulregister\cite7
\soulregister\eqref7
\soulregister\ref7
\sethlcolor{cyan}

\begin{document}
\title{Bessel Vortices in Spin-Orbit Coupled Spin-1 Bose-Einstein Condensates}
\author{Jun-Zhu Li, Huan-Bo Luo, and Lu Li}
\email{llz@sxu.edu.cn}
\affiliation{Institute of Theoretical Physics and Department of Physics, State Key
Laboratory of Quantum Optics and Quantum Optics Devices, Collaborative
Innovation Center of Extreme Optics, Shanxi University, Taiyuan 030006, China}


\begin{abstract}
We investigate the stationary vortex solutions in two-dimensional (2D) Rashba
spin-orbit (SO) coupled spin-1 Bose-Einstein condensate (BEC). By introducing
the generalized momentum operator, the linear version of the system can be
solved exactly and its solutions are a set of the Bessel vortices. Based on
the linear version solutions, the stationary vortex solutions of the full
nonlinear system are constructed and determined entirely by the variational
approximation. The results show that the variational results are in good
agreement with the numerical ones. By means of the variational results, the
vortex ground state phase-transition between the stationary vortex solutions,
stability, and the unit Bloch vector textures are discussed in detail. The
results have the potential to be realized in experiment.

\end{abstract}
\maketitle

\section{Introduction}

Atomic Bose-Einstein condensates (BECs), as an extremely clean quantum system
with full controllability, have been used to emulate various effects from
condensed matter systems~\cite{RepProgPhys.75.082401}. A well-know example is
the spin-orbit (SO) coupling, which plays an important role in spin Hall
effects~\cite{RevModPhys.82.1959}, topological
insulators~\cite{RevModPhys.82.3045}, spintronic
devices~\cite{RevModPhys.76.323}, etc. The last decade has witnessed the
experimental realization of the SO coupling in BECs from one
dimension~\cite{nature09887,PRL108_235301} to two
dimension~\cite{Science.354.83}, see reviews of the experimental and
theoretical findings in Refs.~\cite{Spielman,Galitski,Ohberg,Zhai}. At the
same time, many remarkable characteristics, such as
vortices~\cite{PhysRevLett.107.200401,Kawakami,Drummond,Sakaguchi},
skyrmions~\cite{PhysRevLett.109.015301} and
solitons~\cite{PhysRevLett.110.264101,1D sol 2,1D sol 3,1D sol
4,PhysRevE.89.032920,PhysRevE.94.032202,Cardoso,Lobanov,2D SOC gap sol
Raymond,SOC 2D gap sol Hidetsugu,low-dim SOC,Han Pu 3D}, have been predicted
in the SO coupled BECs theoretically, see also review~\cite{SOC-sol-review}.
Also, the SO coupling, as a basic effect, takes also part in the study of
chiral supersolid~\cite{PhysRevLett.121.030404} and polariton topological
insulator~\cite{PhysRevLett.122.083902}.

Analytical solutions always play a very important role in understanding the
system, explaining various phenomena and making predictions under certain
conditions. Generally, in the framework of the variational approximation (VA), Gaussian ansatz is 
presupposed in studies of the SO coupled spin-1 BEC~\cite{PhysRevA.95.013608,PhysRevA.103.L011301}. Although 
the variational method under Gaussian ansatz is easy to implement, its 
accuracy and application scope are limited. Recently, based on the Bessel vortex solutions of the linear
version of the system, Josephson oscillations of chirality and Bessel vortices
in the SO coupled spin-1/2 BECs have been studied~\cite{Boris,Luohuanbo}.
Compared with Gaussian ansatz, these investigations provide a more accurate way to study the vortex dynamics of the nonlinear system.
Inspired by them, we will consider the analytical vortex solutions of two-dimensional (2D) SO coupled spin-1 BECs
and find some different properties between pseudospin-1/2 BECs and spin-1 BECs, including the occurrence
of polar state in SO coupled spin-1 BECs and their topological properties.

In this paper, we will first show the analytical solutions of the linear
version of spin-1 BECs with the SO coupling by means of Bessel function. It is
found that the Bessel vortices are the fundamental solution of the linear
version of spin-1 BECs with the SO coupling. In the presence of attractive
contact interactions, the Bessel vortex solutions of the linear equation are
modified by multiplying a truncation function and the vortex states of the
full nonlinear system are determined by means of the variational approximation. 
All the variational solutions are perfectly matched with the numerical
results. With the help of the variational solutions, the phase-transition between the vortex ground states, stability
and topological properties are also discussed.

The rest of this paper is structured as follows. In Sec. II, the theoretical
model is introduced. In Sec. III, the linear solution is constructed by means
of Bessel function. Based on the linear vortex solutions, the nonlinear vortex
solutions is constructed by means of the variational approximation in Sec. IV.
At the same time, the vortex ground state of the system, stability, and the
unit Bloch vector textures are discussed in detail. Finally, the main results
of the paper are summarized in Sec. V.

\section{Model and its reductions}

We consider a SO-coupled spin-1 BEC with attractive contact interaction in 2D
space. The spinor wave function, $\Psi=(\Psi_{+1},\Psi_{0},\Psi_{-1})^{T}$, of
this system can be governed by Gross-Pitaevskii (GP) equations in the
dimensionless form
\begin{equation}
i\partial_{t}\Psi= [ -\nabla_{\bot}^{2}/2 + i\beta(F_{y}\partial_{x} -
F_{x}\partial_{y}) + c_{0}\rho+ c_{2}\rho\mathbf{S}\cdot\mathbf{F} ]
\Psi,\label{main}%
\end{equation}
where $\beta$ is the SO coupling strength and $\mathbf{F}=(F_{x},F_{y},F_{z})$
is a vector of spin-1 matrices with
%\footnotesize%
\begin{equation}%
\begin{split}
F_{x} & =\frac{1}{\sqrt{2}}
\begin{pmatrix}
0 & 1 & 0\\
1 & 0 & 1\\
0 & 1 & 0
\end{pmatrix}
, F_{y}=\frac{1}{\sqrt{2}}
\begin{pmatrix}
0 & -i & 0\\
i & 0 & -i\\
0 & i & 0
\end{pmatrix},\\
F_{z} & =
\begin{pmatrix}
1 & 0 & 0\\
0 & 0 & 0\\
0 & 0 & -1
\end{pmatrix}
.
\end{split}
\label{F}%
\end{equation}
%\normalsize
The coefficients $c_{0}<0$ and $c_{2}<0$ are the strengths of the
particle-attraction and spin-attraction contact interactions, respectively.
$\rho=\Psi^{\dag}\Psi$ is the particle density and $\mathbf{S}=\Psi^{\dag
}\mathbf{F}\Psi/\rho$ is the local spin. Below, we fix $\beta= 1$ by means of rescaling.

Stationary solutions of Eq. (\ref{main}) with chemical potential $\mu$ are
sought for in the usual form
\begin{equation}
\Psi(x,y,t)=\psi(x,y)\exp(-i\mu t)\label{P}%
\end{equation}
with stationary functions $\psi(x, y)=(\psi_{+1},\psi_{0},\psi_{-1})^{T}$
satisfying the following equation
\begin{equation}
\mu\psi= [ -\nabla_{\bot}^{2}/2+ i(F_{y}\partial_{x} - F_{x}\partial_{y}) +
c_{0}\rho+ c_{2}\rho\mathbf{S}\cdot\mathbf{F} ] \psi.\label{stationary}%
\end{equation}
In this paper, we will seek the stationary vortex solutions for
Eq.~\eqref{stationary} in the form
\begin{equation}%
\begin{split}
\psi_{1}(x,y) & =e^{-i(m+1)\theta}R_{1}(r),\\
\psi_{0}(x,y) & =e^{-im\theta}R_{0}(r),\\
\psi_{-1}(x,y) & =e^{-i(m-1)\theta}R_{-1}(r),
\end{split}
\label{vortex solutions}%
\end{equation}
where $(r,\theta)$ are the polar coordinates, $m$ is an integer winding
number, and $R_{1,0,-1}(r)$ are three radial wave functions. Substituting
Eq.~\eqref{vortex solutions} into Eq.~\eqref{stationary} yields%
\begin{equation}%
\begin{split}
\mu R_{\pm1}\! =\!  & - \frac{1}{2}\left[ \partial_{r}^{2}+\frac{1}{r}%
\partial_{r}-\frac{(m\pm1)^{2}}{r^{2}}\right] R_{\pm1}\\
&  \!+ \!\!\frac{1}{\sqrt{2}}\left( \pm\partial_{r}\!-\!\frac{m}{r}\right)
\!\!R_{0}\!+\! c_{0}\!(R_{1}^{2}\!+\!R_{0}^{2}\!+\!R_{-1}^{2})R_{\pm1}\\
&  \pm c_{2}(R_{1}^{2}-R_{-1}^{2})R_{\pm1} + c_{2}R_{0}^{2}(R_{\pm1}+R_{\mp
1}),\\
\mu R_{0}\! =\!  & - \frac{1}{2}\left( \partial_{r}^{2}+\frac{1}{r}%
\partial_{r}-\frac{m^{2}}{r^{2}}\right) R_{0}\\
&  \!- \!\frac{1}{\sqrt{2}}\!\left[ \!\left( \!\partial_{r}\!\!+\!\!\frac
{m+1}{r}\!\right) \!\!R_{+1} \!-\!\left( \!\partial_{r}\!\!-\!\!\frac{m-1}%
{r}\!\right) \!\!R_{-1}\!\right] \\
&  + c_{0}(R_{1}^{2}\!+\!R_{0}^{2}\!+\!R_{-1}^{2})R_{0}\! +\! c_{2}%
(R_{1}\!+\!R_{-1})^{2}R_{0}.
\end{split}
\label{radial}%
\end{equation}
Thus, we can obtain the stationary vortex solutions by solving numerically Eq.~\eqref{radial}.

\section{Exact vortex states of the linear version}

In the section, we will consider the stationary vortex solutions of the linear
version of Eq.~\eqref{stationary}, i.e. $\mu\psi=\hat{H}\psi$, with
Hamiltonian
\begin{equation}
\hat{H}=-\nabla_{\bot}^{2}/2+i(F_{y}\partial_{x} - F_{x}\partial_{y}).
\end{equation}

First, we introduce a generalized momentum operator
\begin{equation}
\hat{P}=iF_{x}\partial_{y} - iF_{y}\partial_{x},\label{PP}%
\end{equation}
whose eigenvalue equation, $k\psi=\hat{P}\psi$ with real $k$, admits a set of
exact eigen-states in the form~\eqref{vortex solutions} with
\begin{equation}
R_{1}=\frac{J_{m+1}(kr)}{\sqrt{2}} , R_{0}=J_{m}(kr) , R_{-1}=\frac
{J_{m-1}(kr)}{\sqrt{2}}, \label{psi_1}%
\end{equation}
where $J_{m}$ denotes the Bessel function with integer number $m$ and $k$ is
the radial momentum.

Next, we turn to the eigenvalue problem of the linear Hamiltonian $\hat{H}$.
It can be directly shown that the Hamiltonian $\hat{H}$ commutes with the
generalized momentum operator $\hat{P}$, i.e. $[\hat{H},\hat{P}]=0$. Thus the
eigen-state $\psi$ given by Eq.~\eqref{vortex solutions} with the
expression~\eqref{psi_1} is also the eigen-state of $\hat{H}$. Also, the
Hamiltonian can be written as $\hat{H}=\hat{P}^{2}/2-\hat{B}/4-\hat{P}$, where
$\hat{B}=(F_{z}^{2}+F_{x}^{2}-F_{y}^{2})\partial_{xx}+(F_{z}^{2}-F_{x}%
^{2}+F_{y}^{2})\partial_{yy}+2(F_{x}F_{y}+F_{y}F_{x})\partial_{xy}$ meeting
$\hat{B}\psi=0$, and so the corresponding chemical potential is
\begin{equation}
\mu=k^{2}/2-k, \label{mu}%
\end{equation}
where the two terms represent the kinetic energy and SO-coupling energy. The
chemical potential attains its minimum, $\mu_{\text{min}}=-1/2$ at $k = 1$.
Thus, the solution~\eqref{vortex solutions} with Eq.~\eqref{psi_1} presents a
set of the Bessel vortices with winding numbers $-(m+1)$, $-m$, and $-(m-1)$
in the $\psi_{1}$, $\psi_{0}$, and $\psi_{-1}$ components, respectively. The
similar exact solutions of the SO-coupled binary linear GP equations were
recently reported in Ref.~\cite{Boris,Luohuanbo}. Note that all the vortex
states are degenerate with respect to the excitation number $m$, as $\mu$,
given by Eq.~\eqref{mu}, does not depend on $m$.

Naturally, the norm integral for this linear state in the free space diverges
as
\begin{equation}%
\begin{split}
N &  \equiv N_{1}+N_{0}+N_{-1}\\
&  =\lim_{R\rightarrow\infty}\left\{  2\pi\int_{0}^{R}[R_{1}^{2}+R_{0}%
^{2}+R_{-1}^{2}]rdr\right\}  \simeq4R,
\end{split}
\label{N}%
\end{equation}
while the ratio of the norms of the three components is finite:
\begin{equation}
N_{1}:N_{0}:N_{-1}=1:2:1.\label{N1N2N3}%
\end{equation}
The linear Bessel vortices indicate that particles propagate radially with
momentum $k$. The fact that the norm diverges means that such states require
an infinite number of particles and therefore cannot be realized in real
world. As shown below, taking into account the self-attractive nonlinearity in
Eq.~\eqref{stationary} makes it possible to replace the vortex states by
similar ones, but with a finite norm.

\section{Construction of nonlinear vortex states}

In the following, based on the linear vortex solutions mentioned above and by
means of the variational approximation (VA), we will construct the stationary
vortex solutions of full nonlinear Eq.~\eqref{stationary}. In this process, a
reasonable assumption on ansatz is crucial.

We first consider the asymptotic expression of the vortex solutions for
Eq.~\eqref{stationary}, which should be of the form
\begin{equation}%
\begin{split}
R_{1}  & \mathop{\approx}_{r\rightarrow\infty} \frac{C}{\sqrt{2r}}%
e^{-\sqrt{-2\mu-1}r}\sin\left[ r-\frac{\pi}{2}\left( m+\frac{1}{2}\right)
\right] ,\\
R_{0}  & \mathop{\approx}_{r\rightarrow\infty} \frac{C}{\sqrt{r}}%
e^{-\sqrt{-2\mu-1}r}\cos\left[ r-\frac{\pi}{2}\left( m+\frac{1}{2}\right)
\right] ,\\
R_{-1}  & \mathop{\approx}_{r\rightarrow\infty} -\frac{C}{\sqrt{2r}}%
e^{-\sqrt{-2\mu-1}r}\sin\left[ r-\frac{\pi}{2}\left( m+\frac{1}{2}\right)
\right] ,\\
\end{split}
\label{asymptotic}%
\end{equation}
where $C$ is a constant. Indeed, it can be verified that the asymptotic
expression~\eqref{asymptotic} is valid by substituting it into
Eq.~\eqref{radial} and ignoring infinitesimal of higher order. Thus, the
localized states exist at value of the chemical potential $\mu<-1/2$ in the
nonlinear regime.

The self-focusing nonlinearity chops off the slowly decaying tails of the
Bessel wave function, which make its integral norm diverging. To take this
effect into account in the framework of the VA, we adopt an ansatz based on
the wave function~\eqref{psi_1}, multiplied by truncation factor $A
\text{sech}(ar)$, with amplitude $A$ and inverse width $a$:
\begin{equation}%
\begin{split}
R_{1}(r)= &  \frac{A}{\sqrt{2}} \text{sech}(ar)J_{m+1}(kr),\\
R_{0}(r)= &  A \text{sech}(ar)J_{m}(kr),\\
R_{-1}(r)= &  \frac{A}{\sqrt{2}}\text{sech}(ar)J_{m-1}(kr),
\end{split}
\label{ansatz}%
\end{equation}
where $k$ is still the radial momentum because it can be obtained by
evaluating expectation value of the generalized momentum operator~\eqref{PP}
in VA function $\psi$ given by Eqs.~\eqref{vortex solutions}
and~\eqref{ansatz}, i.e., $\iint\psi^{\dagger}\hat{P}\psi dxdy/\iint%
\psi^{\dagger}\psi dxdy = k$. The truncation factor $\text{sech}(ar)$ is
chosen here and it is easy to prove that the ansatz~\eqref{ansatz} is
compatible with the asymptotic expression~\eqref{asymptotic} at $k=1$ and
$\mu=-(a^{2}+1)/2$. One can see that $-a^{2}/2$ is the energy shift from
linear version chemical potential $\mu=-1/2$ in the nonlinear case. The
amplitude $A$ can be determined, as a function of $a$ and $k$, by imposing the
normalization condition, $N =\iint\psi^{\dagger}\psi dxdy = 1$.

\begin{figure}[ptb]
\centering
%Requires \usepackage{graphicx}
\includegraphics[width=3.3in]{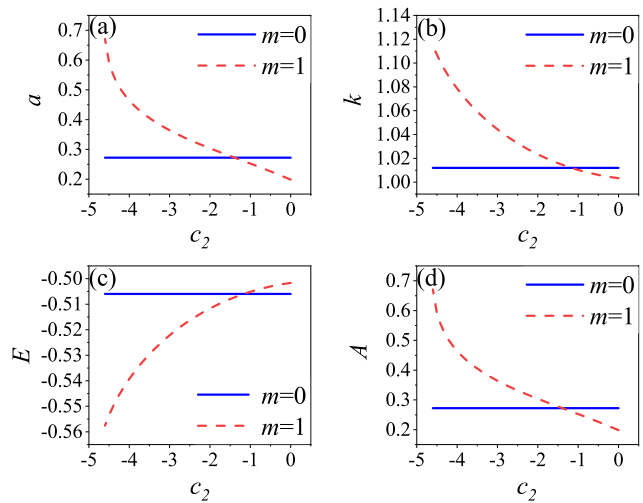}\caption{(Color online) (a) Inverse
width $a$, (b) momentum $k$, (c) total energy $E$, and (d) real amplitude $A$
as a function of interaction strength $c_{2}$ at $c_{0}=-1.5$. Here, the blue
and red curve correspond to the state with $m=0$ and $m=1$, respectively.}%
\label{figure1}%
\end{figure}

In order to determine the parameters $a$ and $k$, we need to minimize the
total energy
\begin{equation}%
\begin{split}
E(a,\!k)\! =\frac{1}{2}\! \iint & \left[ -\psi^{\dag}\nabla_{\bot}^{2}\psi+
2i\psi^{\dag}\left( F_{y}\partial_{x}\!-\!F_{x}\partial_{y}\right) \psi\right.
\\
& \left.  + c_{0}n^{2} + c_{2}n^{2}|\mathbf{S}|^{2}\right]  dxdy.
\end{split}
\label{E}%
\end{equation}
The expression of the total energy $E(a,k)$ can be expended by the
substitution of Eqs.~\eqref{vortex solutions} and~\eqref{ansatz} into
Eq.~\eqref{E} in the polar coordinates, and its minimization can be
numerically implemented by means of the simplex search
method~\cite{SIAM.J.Optim.9.1}. This method differs from that used in Refs.
\cite{PhysRevA.95.013608, PhysRevA.103.L011301}, in which the total energy was
exactly calculated by using Gaussian ansatz. Fig.~\ref{figure1} presents the
corresponding results for the states with $m=0,1$, varying values of $c_{2}$
at $c_{0} = -1.5$. For the state with $m=0$, the parameters $a$ and $k$ remain
unchanged as $c_{2}$ varies from $-5$ to $0$, resulting in the total energy
$E$ and amplitude $A$ unchange too. The result can be interpreted by the fact
that $\mathbf{S}=0$ for the state with $m=0$. For the state with $m=1$, it is
found that with the increasing of $|c_{2}|$, the parameters $a$ and $k$
increase and the total energy $E$ decreases, where the inverse width $a$
increases sharply as $c_{2}\rightarrow-4.6$ until it becomes infinity at
$c_{2}=-4.6$, as shown in Fig.~\ref{figure1}(a), which means that the collapse
occurs~\cite{Phys.Rep.303.259,Phys.Rep.507.43}. Also, one can see from
Fig.~\ref{figure1}(c) that the total energy of the state with $m=0$ is less
than that of the state with $m=1$ as $c_{2}>-1.2$, and the opposite is true as
$c_{2}<-1.2$. Thus, for given $c_{0}=-1.5$, $c_{2}=-1.2$ provides a vortex
ground state phase-transition point, at which the vortex ground state is
transformed into the state with $m=1$ from the state with $m=0$.

\begin{figure}[ptb]
\centering
%Requires \usepackage{graphicx}
\includegraphics[width=3.0in]{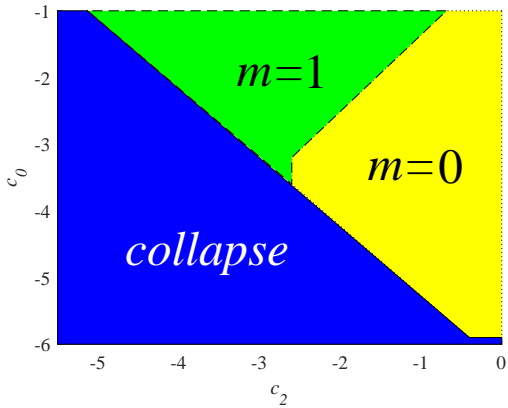}\caption{(Color online) Diagram of the
vortex ground state described by the states with $m=0$ and $m=1$ on
$(c_{0},c_{2})-$plane.}%
\label{figure2}%
\end{figure}

It should be emphasized that Fig.~\ref{figure1} presents only the results of
$c_{0}=-1.5$. In general, the diagram of the vortex ground state described by
the state with $m=0$ and the state with $m=1$ on $(c_{0},c_{2})-$plane is
shown in Fig.~\ref{figure2}. From it, one can see that for given a
particle-attractive interaction strength $c_{0}>-3.6$, increasing the
spin-attractive interaction strength $|c_{2}|$ will cause the vortex ground
state phase-transition from the state with $m=0$ to the state with $m=1$ until
collapse. While as $-5.9<c_{0}<-3.6$, the state with $m=0$, as the vortex
ground state, collapses directly with the increase of $|c_{2}|$. As
$c_{0}<-5.9$, only collapse occurs for all of $c_{2}$%
~\cite{Phys.Rep.303.259,Phys.Rep.507.43,PhysRevA.91.043604}. Furthermore, one find that
the threshold of $|c_0|$, above which the collapse occurs, will
decrease with increasing of $|c_2|$. This means that the spin-attraction
interaction increases the centripetal velocity of the particles, i.e., the attraction-induced velocity~\cite{PhysRevA.91.043604}.

\begin{figure}[ptb]
\centering
%Requires \usepackage{graphicx}
\includegraphics[width=3.4in]{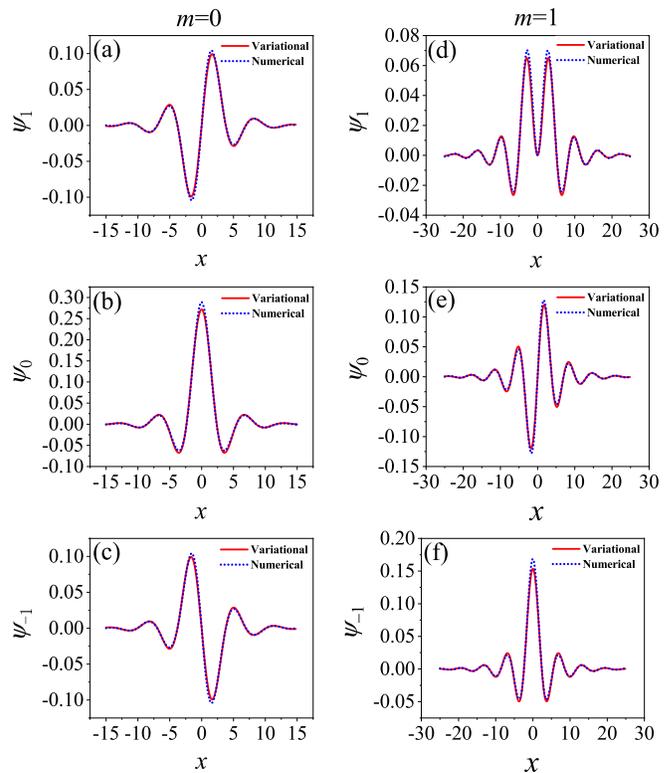}\caption{(Color online) Distributions
of the vortex ground state wave function at $y=0$, where (a)-(c) $c_{0}=-1.5$,
$c_{2}=-0.5$, and the blue dotted curves correspond to the state with $m=0$,
and (d)-(f) $c_{0}=c_{2}=-1$ and the blue dotted curves are to the state with
$m=1$.}%
\label{figure3}%
\end{figure}

In order to verify the correctness of the variational results, we solved
Eq.~\eqref{main} numerically by employing the norm-preserving imaginary time
propagation method. As typical examples, Fig.~\ref{figure3} presents the
distributions of the vortex ground state wave function at $y=0$ for two
different sets of parameters $c_{0}=-1.5,c_{2}=-0.5$ and $c_{0}=c_{2}=-1$,
which correspond to the state with $m=0$ and the state with $m=1$ given by
Eq.~\eqref{vortex solutions} with the ansatz~\eqref{ansatz}, respectively.
Here, the parameters in Eq.~\eqref{ansatz} calculated by VA are $a=0.2877$,
$k=1.0120$, $A=0.2723$ for the state with $m=0$ and $a=0.1816$, $k=1.0041$,
$A=0.2173$ for the state with $m=1$, respectively. One can see that numerical
wave functions are nearly identical to the VA counterparts, except for a
slight mismatch at peaks and valleys. The reasons can be summarized in three
aspects. Firstly, the variational solution~\eqref{ansatz} can be reduced to
the linear exact solution~\eqref{psi_1} at $a=0$, and so is an exact solution
in linear case ($c_{0,2}=0$). Secondly, in nonlinear case, since the
truncation function is determined according to the asymptotic
expression~\eqref{asymptotic}, the tail of the wave function can always be
approximated very well by the variational solution. Finally, the nonlinear
effect is positively correlated with the nonlinear coefficient $c_{0,2}$ and
the amplitude of the wave functions $\psi_{1,0,-1}$. This is the reason for
the mismatch between numerical solution and variational solution at the peaks
and valleys. In the case of nonlinear coefficients $|c_{0,2}|<6$, the mismatch
is kept within the tolerable range.

It is relevant to mention that, the variational method under the Gaussian
ansatz for solving the similar system has been reported in
Ref.~\cite{PhysRevA.95.013608}. It is necessary to compare Gaussian ansatz
with Bessel ansatz in terms of easy of implementation, accuracy and scope of
application. The variational method under the Gaussian ansatz is easier to
implement, in which total energy and variational parameters for the lowest
energy can be obtained analytically, while they can only be obtained
numerically under the Bessel ansatz. However, Bessel ansatz is more accurate
and has wilder application scope than Gaussian ansatz. The Bessel ansatz tends
to become linear exact solutions in the case of weak nonlinearity and the
truncation function is used to correct the deformation caused by nonlinearity.
These make the Bessel ansatz has a high accuracy in both strong and weak
nonlinearity. On the contrary, the Gaussian ansatz lacks the oscillatory
property (the property of linear exact solution or Bessel function), which
makes it only applicable to the case of relatively strong nonlinearity
($|c_{0}|>4$) and will fail in the case of weak nonlinearity ($|c_{0,2}|<2$).
Also, an important application of Bessel ansatz is to give the phase diagram
of the ground state vortex (see Fig.~\ref{figure2}), which is impossible for
Gaussian ansatz due to the limitation of the application scope. In conclusion,
the results given by the VA under Bessel ansatz can be better used to study
the dynamics and topological properties of the vortex ground state.

\begin{figure}[ptb]
\centering
%Requires \usepackage{graphicx}
\includegraphics[width=3.4in]{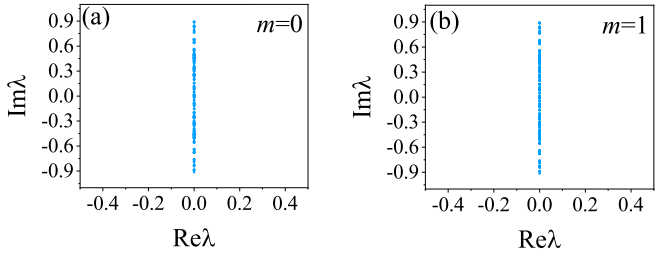}\caption{(Color online) Eigen spectra
of the linear stability analysis. (a) $c_{0}=-1.5$ and $c_{2}=-0.5$, and (b)
$c_{0}=c_{2}=-1$. }%
\label{figure4}%
\end{figure}

\begin{figure}[ptb]
\centering
%Requires \usepackage{graphicx}
\includegraphics[width=3.4in]{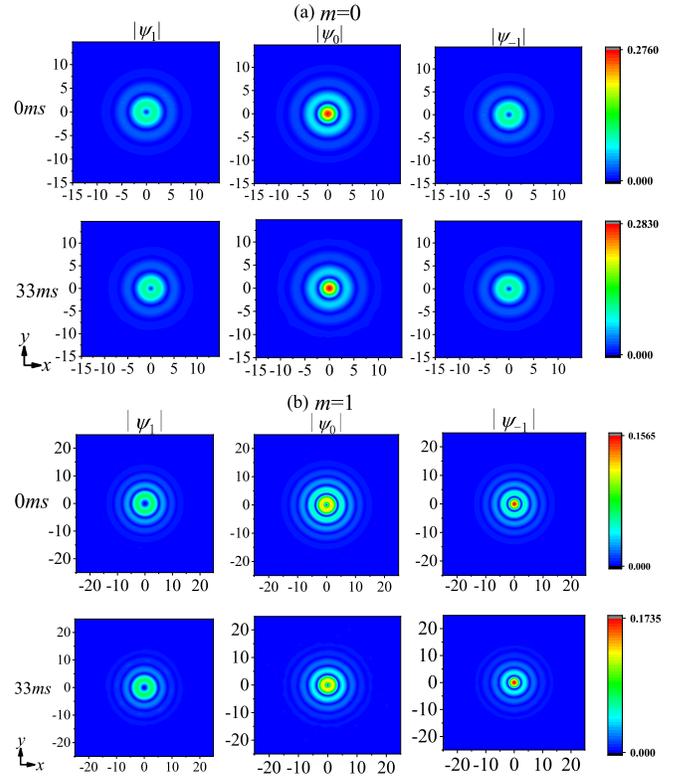}\caption{(Color online) Numerical
evolutions of the perturbed vortex ground state with $m=0$ and $m=1$ under the
initial random perturbation whose maximal value is 0.03. (a) $c_{0}=-1.5$ and
$c_{2}=-0.5$, and (b) $c_{0}=c_{2}=-1$. Here, the first row is the initial
inputs and the second row is the outputs at $t=33$ms in (a) and (b),
respectively. }%
\label{figure5}%
\end{figure}

Now, by means of the results of the VA under Bessel ansatz, we discuss the
stability of the vortex ground state by employing linear stability analysis
and direct simulation, respectively. The linear stability analysis can be
performed by adding a perturbation to the stationary vortex solution
$\psi(x,y)$ for Eq.~\eqref{stationary}
\begin{equation}
\Phi(x,\!y,\!t)\!=\!e^{\!-i\mu t}\!\!\left[ \psi(x,\!y)\!+\!w(x,\!y)e^{\lambda
t}\!+\!v^{*}(x,\!y)e^{\!\lambda^{*}t}\right] ,\label{perturbation}%
\end{equation}
where $w(x,y)$ and $v(x,y)$ are small perturbation vectors, $\lambda$ is the
eigenvalue, and the asterisk stands for the complex conjugation. Substituting
Eq.~\eqref{perturbation} into Eq.~\eqref{main} and linearizing with respect to
the perturbations, we arrive at the linear eigenvalue problem
\begin{equation}
\left(
\begin{matrix}
L_{1} & L_{2}\\
L_{2}^{\ast} & L_{1}^{\ast}%
\end{matrix}
\right)  \left(
\begin{split}
w\\
v
\end{split}
\right)  = \lambda\left(
\begin{split}
w\\
v
\end{split}
\right) ,\label{linear_eigenvalue_problem}%
\end{equation}
where $L_{1}=i\nabla_{\bot}^{2}/2+\beta(F_{y}\partial_{x}-F_{x}\partial
_{y})+i\mu-ic_{0}(\psi^{\dag}\psi+\psi\psi^{\dag})-ic_{2}(\mathbf{F}\psi
\cdot\psi^{\dag}\mathbf{F}^{\dag}+\psi^{\dag}\mathbf{F}\psi\cdot\mathbf{F})$
and $L_{2}=-ic_{0}\psi\psi^{T}-ic_{2}\mathbf{F}\psi\cdot\psi^{T}\mathbf{F}%
^{T}$. Notice that $L_{1}$ and $L_{2}$ are $3\times3$ matrices.
Eq.~\eqref{linear_eigenvalue_problem} can be solved by employing numerical
simulation. The vortex solution $\psi(x,y)$ is linear unstable by decaying or
rising exponentially if $\lambda$ contains real part, otherwise it is linear
stable as complex exponential oscillation with small perturbation does not
influence the stability of $\psi(x,y)$. Fig.~\ref{figure4} presents the eigen
spectra of the linear stability analysis for two different sets of parameters
shown in Fig.~\ref{figure3}. From it, one can see that their eigen spectra
hardly contain the real part, and so the states with $m=0$ and $m=1$, as the
vortex ground state, are linear stable. Furthermore, we demonstrated their
perturbed dynamics by dint of direct simulations of Eq.~\eqref{main}, and the
results are summarized in Fig.~\ref{figure5}. As predicted by linear stability
analysis, the vortex ground state solutions are stable.

The stability of Bessel vortices can be interpreted by the theory described in
Ref~\cite{PhysRevA.91.043604}. The radial momentum $k$ of Bessel vortices, which
corresponds to the anomalous velocity in Ref~\cite{PhysRevA.91.043604},
forms a centrifugal component in the density flux opposite to that arising due to the
attraction between particles and prevents the collapse. Therefore, the stability of
Bessel vortices can be guaranteed by balancing the attraction interaction and the spin-
orbit coupling strength.

The spin-orbit coupling in the BEC Hamiltonian is tantamount to the
Dzyaloshinskii-Moriya interaction, supporting topologically nontrivial spin
textures, i.e., skyrmions~\cite{PhysRevLett.108.185301}. Here we discuss the
topological properties of the vortex ground state with
$m=0$. It is found that the vortex ground state with $m=0$ is a polar state
due to $|\mathbf{S}|^{2}=0$~\cite{PhysRevLett.81.742}. Also, from the eigen
equation $\mathbf{n}\cdot\mathbf{F}\Phi=\delta\Phi$, where $\mathbf{n}%
=(n_{x},n_{y},n_{z})$ is the unit Bloch vector, one can obtain that the
eigenstate with $\delta=0$ is of the form $\Phi_{0}=((-n_{x}+in_{y})/\sqrt
{2},n_{z},(n_{x}+in_{y})/\sqrt{2})^{T}$. Thus, the vortex ground state with
$m=0$ can be written as
\begin{equation}
\psi=\sqrt{\rho}e^{i\vartheta}
\begin{pmatrix}
\frac{-n_{x}+in_{y}}{\sqrt{2}}\\
n_{z}\\
\frac{n_{x}+in_{y}}{\sqrt{2}}%
\end{pmatrix}
, \label{polar state}%
\end{equation}
where $\rho=\psi^{\dag}\psi$ is the particle density,
$\vartheta$ is the superfluid phase~\cite{PhysRevLett.100.180403}. It is also
invariant under simultaneous transformations $\vartheta\rightarrow
\vartheta+\pi$ and $\mathbf{n}\rightarrow-\mathbf{n}$. Thus, the order
parameter manifold for the polar phase can be given by $M=(U(1)\times
S^{2})/\mathbb{Z}_{2}$, where $U(1)$ denotes the manifold of the superfluid
phase $\vartheta$, and $S^{2}$ is 2D sphere whose point specifies the
direction of $\mathbf{n}$.

\begin{figure}[ptb]
\centering
%Requires \usepackage{graphicx}
\vspace{9mm} \includegraphics[width=2.8in]{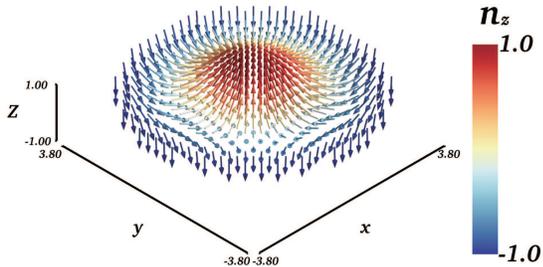}\caption{(Color online)
Unit Bloch vector texture of the vortex ground state with $m=0$ in the region
of $r\leq r_{1}=3.8$, where the system parameters are $c_{0}=-1.5$ and
$c_{2}=-0.5$.}%
\label{figure6}%
\end{figure}

By comparing Eq.~\eqref{polar state} and Eq.~\eqref{vortex solutions} with the
ansatz~\eqref{ansatz} with $m=0$, it can be found that $n_{x}=\sin{\Phi
(r)}\cos{\theta}$, $n_{y}=\sin{\Phi(r)}\sin{\theta}$, $n_{z}=\cos{\Phi(r)}$,
where $\Phi(r)$ is the polar angle of $\mathbf{n}$ and $\sin{\Phi(r)}%
=-J_{1}(kr)/\sqrt{J_{1}^{2}(kr)+J_{0}^{2}(kr)}$, $\cos{\Phi(r)}=J_{0}%
(kr)/\sqrt{J_{1}^{2}(kr)+J_{0}^{2}(kr)}$. From those expressions, one can
obtain that $\mathbf{n}(r_{l})=(0,0,(-1)^{l})$, where $r_{0}=0$ and $r_{l}$ is
the $l$-th root of $J_{1}(kr)=0$ with $l=1,2,\dots$. This implies that the
unit Bloch vector is pointing in the positive direction of $z$ at the origin
point, and as $r=r_{1}$ it is pointing in the opposite direction of $z$.
The domain covered by the vertex of the unit Bloch vector from $r=r_0$ to $r=r_1$ forms an enclosed area
and can be compactified into a $S^{2}$. A given mapping $\mathbf{n}:S^{2}\rightarrow S^{2}$ determines
the skyrmion topological number $Q$. The skyrmion topological number counts the number of times that
the $S^{2}$ for $\mathbf{n}$ is covered and can be calculated in a finite area where the skyrmion is restricted.
For the simplicity, we only consider the case of $Q=1$, i.e., the skyrmion restricted in the area with the boundary $r=r_1$. Thus the skyrmion can be
classified by the second homotopy group $\pi_{2}(S^{2})=\mathbb{Z}$ and characterized by topological number as following formula
\begin{equation}%
\begin{split}
Q =  &  \frac{1}{4\pi}\iint_{\Sigma}rdrd\theta\\
&  \mathbf{n}\!\!\cdot\!\!\left[ \! \left( \!\cos{\theta}\partial
_{r}\!\!-\!\!\frac{\sin{\theta}}{r}\partial_{\theta}\!\right) \mathbf{n}
\!\times\!\!\left( \!\sin{\theta}\partial_{r}\!\!+\!\!\frac{\cos{\theta}}%
{r}\partial_{\theta}\!\right) \mathbf{n} \!\right] ,
\end{split}
\label{topological number}%
\end{equation}
where the integral domain $\Sigma:0\leq r\leq r_{1}$, $0\leq\theta<2\pi$. The
skyrmion topological number of the vortex ground state with $m=0$ in the
region of $r\leq r_{1}$ is given by $Q=-0.5\cos\Phi(r)|_{r=0}^{r=r_{1}%
}=-0.5n_{z}|_{n_{z}=1}^{n_{z}=-1}=1$. Fig.~\ref{figure6} shows the unit Bloch
vector texture of the vortex ground state with $m=0$ in the region of $r\leq
r_{1}=3.8$. From it, one can see that the unit Bloch vector surrounds the
sphere $S^{2}$ once, forming a skyrmion with $Q=1$.

\section{Conclusions}

In summary, we investigated the stationary vortex solutions in 2D Rashba SO
coupled spin-1 BEC with attractive contact interaction. The linear version of
the system can be solved exactly by introducing the generalized momentum
operator. The linear version solution is a sets of Bessel vortices. Based on
the Bessel vortices and by means of variational approximation, we also given
out the solutions of full nonlinear system. The results have shown that the
variational results are in good agreement with the numerical results, and can
stably evolve, which can meet the requirements of long-time observation in
experiment. We also investigated the vortex ground state phase-transition
between the eigen-states with $m=0$ and $m=1$ and the unit Bloch vector
texture of the vortex ground state with $m=0$. It has found that the latter
can form a skyrmion structure with topological number $Q=1$.

\section*{ACKNOWLEDGMENTS}

This research was supported by 111 project (Grant No. D18001), and the Hundred
Talent Program of the Shanxi Province (2018).

\end{document}